# Enhancement and suppression of surface plasmon resonance in Ag aggregate by optical gain and absorption in surrounding dielectric medium

M. A. Noginov[1], G. Zhu[1], V. M. Shalaev[2], V. P. Drachev[2], M. Bahoura[1], J. Adegoke[1], C. Small[1], B. A. Ritzo[3]

[1] *Center for Materials Research, Norfolk State University, Norfolk, VA 23504*
*mnoginov@nsu.edu*

[2] *School of Electrical & Computer Engineering, Purdue University, West Lafayette, IN 47907*

[3] *Summer Research Program, Center for Materials Research, Norfolk State University, Norfolk, VA 23504*

We have observed the compensation of loss in metal by gain in interfacing dielectric in the mixture of aggregated silver nanoparticles and rhodamine 6G dye. The demonstrated six-fold enhancement of the Rayleigh scattering is the evidence of the increase of the quality factor of the surface plasmon (SP) resonance. The reported experimental observation paves the road to many practical applications of nanoplasmonics. We have also predicted and experimentally observed a suppression of the surface SP resonance in metallic nanoparticles embedded in a dielectric host with absorption.

Localized surface plasmon (SP) is an oscillation of free electrons in a metallic particle, whose resonance frequency is the plasma frequency adjusted by the size and, mainly, the shape of the particle. A phenomenon relevant to localized SPs is a surface plasmon polariton (SPP) or a surface electromagnetic wave propagating along the interface between two media possessing permittivities with opposite signs, such as metal-dielectric interface.

Localized plasmons have been found on rough surfaces [1,2], in engineered nanostructures [3-6], as well as in clusters and aggregates of nanoparticles [7-9]. In the spots where local fields are concentrated, both linear and nonlinear optical responses of molecules and atoms are gigantically enhanced. This leads to a number of important applications, the most matured of which is the Surface Enhanced Raman Scattering (SERS) [2]. A number of interesting optical phenomena (such as harmonic generation, SERS, Kerr effect, *etc*.) caused by dramatic field enhancement in hot spots of *fractal aggregates* of silver nanoparticles have been theoretically predicted and experimentally demonstrated by Shalaev and coauthors [10]. Other applications of resonating metallic nanostructures and related phenomena, described by the common name *nanoplasmonics*, include near-field microscopy [11-13], extraordinary transmission of light through arrays of subwavelength holes in metallic films [14], negative index metamaterials [15,16], *etc*.

Most of existing and potential future applications of nanoplasmonics suffer from damping caused by metal absorption. In 1989 Sudarkin and Demkovich suggested to increase the propagation length of SPP by creating the population inversion in the dielectric medium adjacent to the metallic film [17]. Recently, gain-assisted propagation of SPPs at the interface between metal and a dielectric with optical gain has been analyzed theoretically in Refs. [18,19]. The enhancement of SPP (of the order of $10^{-5}$–$10^{-4}$) at the interface between silver film and dielectric medium with optical gain (laser dye) has been recently demonstrated in Ref. [20].

In a similar way, the *localized* SP resonance in metallic nanospheres was predicted to exhibit a singularity when the surrounding dielectric medium has a critical value of optical gain [21]. This singularity, resulting from canceling *both real* and *imaginary* terms in the denominator of the field enhancement factor in a metallic particle $\propto(\varepsilon_d-\varepsilon_m)/(2\varepsilon_d+\varepsilon_m)$ can be evidenced by an increase of the Rayleigh scattering [21] (here $\varepsilon_d$ and $\varepsilon_m$ are complex dielectric constants of dielectric and metal, respectively).

Let us estimate critical gain needed to compensate metal loss of localized SPs. The polarizability (per unit volume) for isolated metallic nanoparticles is given by $\beta=(4\pi)^{-1}[\varepsilon_m-\varepsilon_d]/[\varepsilon_d+p(\varepsilon_m-\varepsilon_d)]$, where $p$ is the depolarization factor (for a sphere, $p$=1/3) [10]. If the dielectric is an active medium with $\varepsilon_d''=-p\varepsilon_m''/(1-p)$, then at the resonance both the real and imaginary parts in the denominator become zero, leading to extremely large local fields limited only by saturation effects [21,22]. Thus, for the gain coefficient needed to compensate loss of *localized* SP we find $\gamma=(2\pi/\lambda_0)\varepsilon_d''\,|\varepsilon_d'|^{-1/2}$ $=(2\pi/\lambda_0)(\Gamma/\omega_p)[p/(1-p)](\varepsilon_0+2\varepsilon_d')^{3/2}\,|\varepsilon_d'|^{-1/2}\sim10^3$ $cm^{-1}$ at $\lambda_0$=0.5 μm (we used the Drude formula $\varepsilon_m=\varepsilon_0-\omega_p^2[\omega(\omega+i\Gamma)]$, $\varepsilon_d$=1.7 and known optical constants from Ref. [23]).

The gain $\gamma \sim 10^2 – 10^3$ $cm^{-1}$ needed to compensate loss of SPP or localized SP is within the limits of semiconducting polymers [24] or laser dyes (highly concentrated, $\sim 10^{-2}$ M [21], or adsorbed onto metallic nanoparticles). In *aggregates* of metallic nanoparticles, critical gain can be even smaller (up to ten times) than in single nanoparticles because of small effective form factors $p$ of chains of nanospheres contributing to the spectrum of the aggregate at the given wavelength [25].

Using the same line of arguments as was used in Ref. [21], one can infer that by embedding metallic nanoparticle in a dielectric medium with absorption, one can further increase the imaginary part of the denominator in the field enhancement factor $\propto (\varepsilon_d - \varepsilon_m)/(2\varepsilon_d + \varepsilon_m)$ and, correspondingly, reduce the quality factor of the plasmon resonance and the peak absorption cross section.

The objective of this work was to demonstrate experimentally both predicted effects, the enhancement of the SP by gain and its suppression by loss in a dielectric medium.

Experimentally, we studied absorption, emission, and Rayleigh scattering in the mixtures of R6G dye (Rhodamine 590 Chloride from Exciton) and aggregate of Ag nanoparticles. The starting solutions of R6G in methanol had concentrations of dye molecules in the range $1 \times 10^{-6}$-$2.1 \times 10^{-4}$M. Poly(vinylpyrrolidone)-passivated silver aggregate suspended in ethanol was prepared according to the procedure described in Ref. [24]. The estimated concentration of Ag nanoparticles in the aggregate was $8.8 \times 10^{13}$ $cm^{-3}$. In some particular experiments, we diluted Ag aggregate before mixing it with dye.

The absorption spectrum of Ag aggregate had one structureless band covering the whole visible range and extending to near-infrared (Fig. 1, trace 8). The major feature in the absorption spectrum of R6G is the band peaking at ≈528 nm (Fig. 1a, trace 1 and inset of Fig. 1a). Experimentally, we added Ag aggregate solution to the dye solution by small amounts and recorded absorption spectra of the mixtures (Fig. 1a, traces 2-7). (In different particular experiments, the 'step' size varied between 1% and 25% of the maximal Ag concentration.) We then scaled the absorption spectrum of pure Ag aggregate to fit each of the spectra of the mixtures at ≤450 nm and calculated the difference spectra {*mixture–aggregate*}. The differential spectra obtained this way, normalized to the concentration of the R6G dye in the mixture, reveal the regular absorption band of R6G dye, at ~0.53 μm, and a much broader new absorption band centered at 0.72-0.75 μm (Fig. 1b). The latter broad band can be due to hybrid states formed by R6G molecules chemisorbed (*via* $Cl^-$) onto silver nanoparticles [27] or due to restructuring of the Ag aggregate in the presence of dye molecules.

As one can see in Fig. 1b, with the increase of Ag aggregate in the mixture, the intensity of the absorption band of R6G (calculated using the procedure described above) decreased, with the rate exceeding the rate of the reduction of the R6G concentration. (According to our previous study [28], the observed reduction of the R6G absorption is due to Ag aggregate but not an aggregate solvent without Ag.)

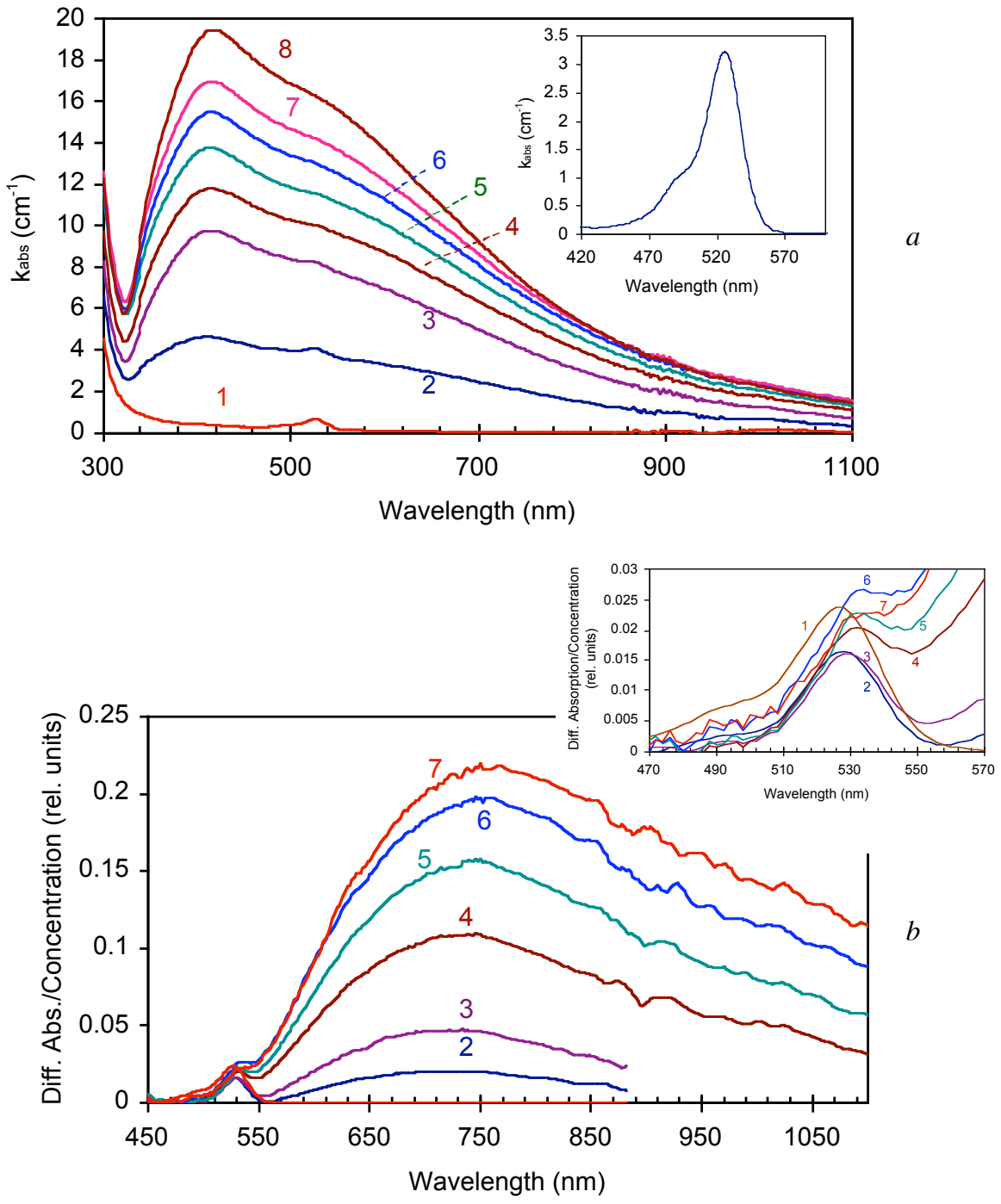

Fig. 1. (a) Trace 1 – Absorption spectrum of the pure R6G dye solution ($2.1 \times 10^{-6}$M); trace 8 – absorption spectra of pure Ag aggregate solution ($8.8 \times 10^{13}$$cm^{-3}$); traces 2-7 – absorption spectra of Ag aggregate-R6G mixtures. The ratio of Ag aggregate solution to R6G solution in the mixture is equal to 27.1/72.9 (2), 57.3/42.7 (3), 66.5/33.5 (4), 74.8/25.2 (5), 80.8/19.2 (6), and 86.8/13.2 (7). Inset: Absorption spectrum of pure R6G dye solution ($1.25 \times 10^{-5}$M). (b) Difference absorption spectra (absorption spectrum of the mixture minus scaled spectrum of the aggregate, normalized to the concentration of R6G dye in the mixture). The traces in Figure b correspond to the spectra with similar numbers in Figure a. Inset: Enlarged fragment of the main frame. Trace 1 corresponds to the pure R6G dye.

Spontaneous emission spectra were studied in the setup schematically shown in the inset of Fig. 2, when the dye or a mixture of dye with Ag aggregate was placed in a 1 mm cuvette. The samples were pumped and the luminescence was collected nearly normally to the surface of the cuvette. We found that while the shape of the emission band was practically unaffected by the presence of Ag aggregate in the mixture, its intensity changed significantly. At the starting concentrations of R6G and Ag aggregate, equal to R6G=1.25x$10^{-5}$ M, Ag=8.7x$10^{12}$ $cm^{-3}$, the emission intensity of the dye (measured in the maximum at ~558 nm) increased up to 21% at the addition of small amounts of aggregated silver nanoparticles (Fig. 2). At the further increase of the concentration of Ag aggregate in the mixture, the emission intensity decreased. However, its relative decrease was much smaller than the relative reduction of the R6G absorption determined as it was described above.

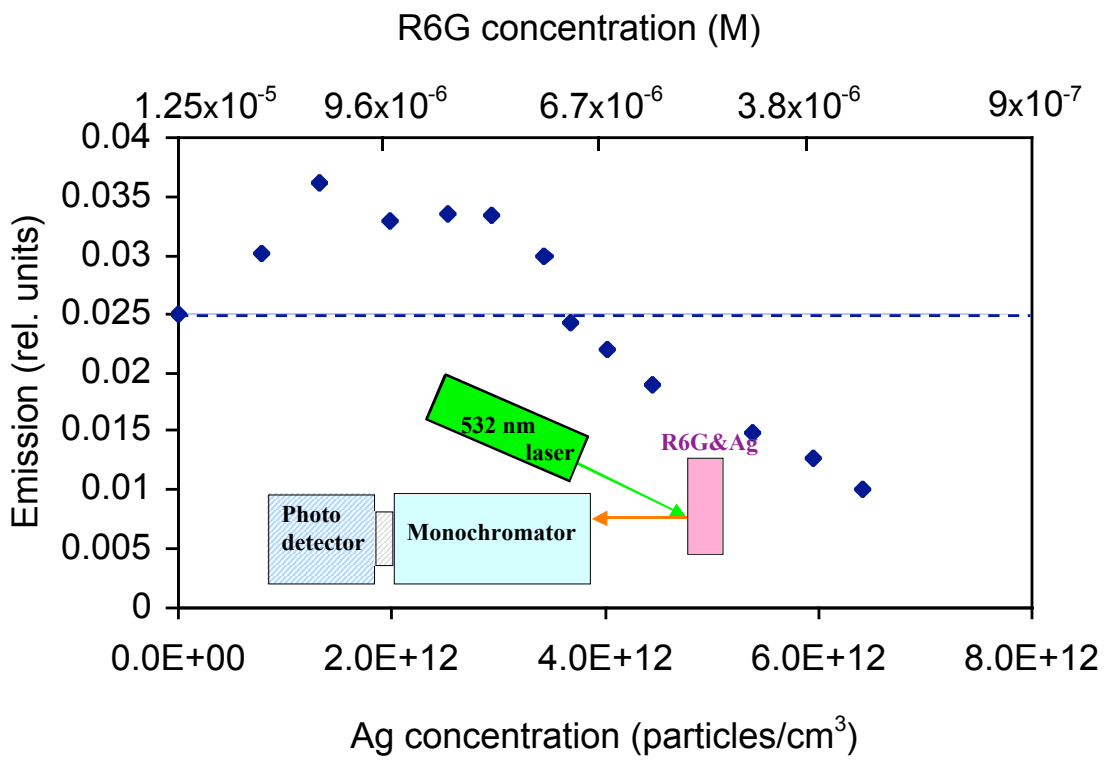

Fig. 2. Emission intensity recorded at the addition of the Ag aggregate to the R6G dye. The starting concentrations of R6G and Ag aggregate are, respectively, 1.25x$10^{-5}$ M and 8.7x$10^{12}$ $cm^{-3}$ The emission intensity corresponds to "as is" detected signal that has not been a subject to any normalization. Inset: Schematic of the experimental setup used at the emission intensity measurement.

According to Refs. [29,30], the quantum yield of spontaneous emission of low concentrated R6G dye is equal to 95%. Thus, the experimentally observed emission enhancement cannot be due to the increase of the quantum yield. An enhancement of emission can be explained by increased absorption of R6G in the presence of Ag aggregate. However, at the first glance, this explanation contradicts with the experimental observation: seeming reduction of the R6G absorption with the increase of Ag aggregate concentration, inset of Fig. 1b.

We argue that the commonly accepted procedure of the decomposition of an absorption spectrum into its components, which we used to treat data of Fig. 1, is not applicable to the mixture of two substances (dye and aggregated metallic nanoparticles) which affect each other, and that the paradox above has a clear physical explanation.

The absorption spectrum of a fractal aggregate is comprised of a continuum of homogenious bands corresponding to metallic nanostructures with different effective form factors. (The homogeneous widths of individual bands are comparable to the characteristic widths of the absorption and emission bands of R6G dye.) Following our prediction of the suppression of the SP resonance by the absorption in a dielectric, we speculate that the absorption band of R6G "burns" a hole in the absorption spectrum of the aggregate in the frequency range corresponding to the absorption of dye. Thus, the conventional method that we used to extract the absorption band of R6G from the absorption spectrum of the mixture was not applicable to our system. Instead, the absorption spectrum of the mixture should be decomposed according to the method schematically shown in Fig. 3. This explanation supports our hypothesis (suppression of SP resonance by loss in dielectric) and resolves the paradox above (an increase of emission intensity at seeming reduction of R6G absorption).

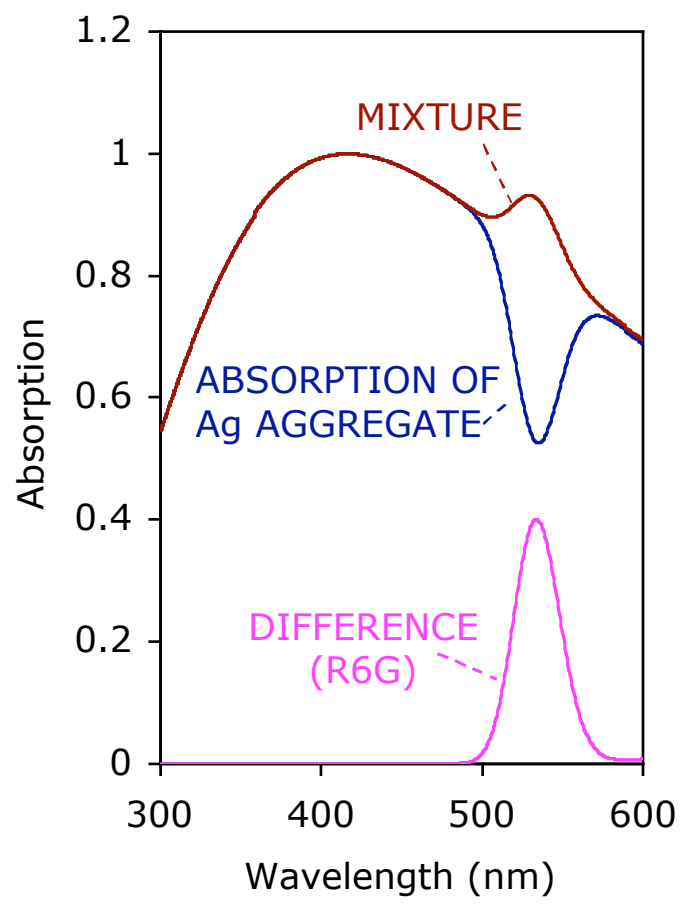

Fig. 3. Schematic of the proposed decomposition of the absorption spectrum of the mixture into the absorption spectra of the components.

As it has been estimated above, the critical gain needed to compensate loss in metallic nanoparticles is of the order of $10^3$ $cm^{-1}$. Obviously, the effective absorption coefficient of the same order of magnitude can cause a significant suppression of the SP resonance. One can estimate that a single molecule of R6G (characterized by the emission cross section ~ 4x$10^{-16}$ $cm^2$) adsorbed onto metallic nanoparticle with the diameter d=10 nm causes the absorption coefficient (per the volume occupied by the nanoparticle) of the order of $10^3$ $cm^{-1}$. Correspondingly, the same *excited* molecule can cause a critical gain of the same order of magnitude. If the number of adsorbed R6G molecules per nanoparticle exceeds one, the effective absorption and gain can be even higher.

Following Lawandy [21], we used an increase of Rayleigh scattering in a pump-probe experiment as an evidence of enhancement of SP by optical gain.

R6G-Ag aggregate mixtures (similar to those used in the absorption measurements) were pumped with a frequency-doubled Q-switched Nd:YAG laser, $\lambda_{pump}$=532 nm, $t_{pump}$≈10 ns. A fraction of the pumping beam was split off and used to pump a simple laser consisting of the cuvette with R6G dye placed between two mirrors, inset of Fig. 4. Since no wavelength selection element was used in the laser cavity, the emission line of the R6G laser (~558 nm) corresponded to the maximum of the gain spectrum of R6G dye in the mixtures studied. The beam of the R6G laser, which was used as a probe in the Rayleigh scattering, was aligned with the pumping beam in the beamsplitter and sent to the sample through a small (0.5 mm) pinhole, inset of Fig. 4. The pump and probe beams were collinear, and their diameters at the pinhole were larger than 0.5 mm.

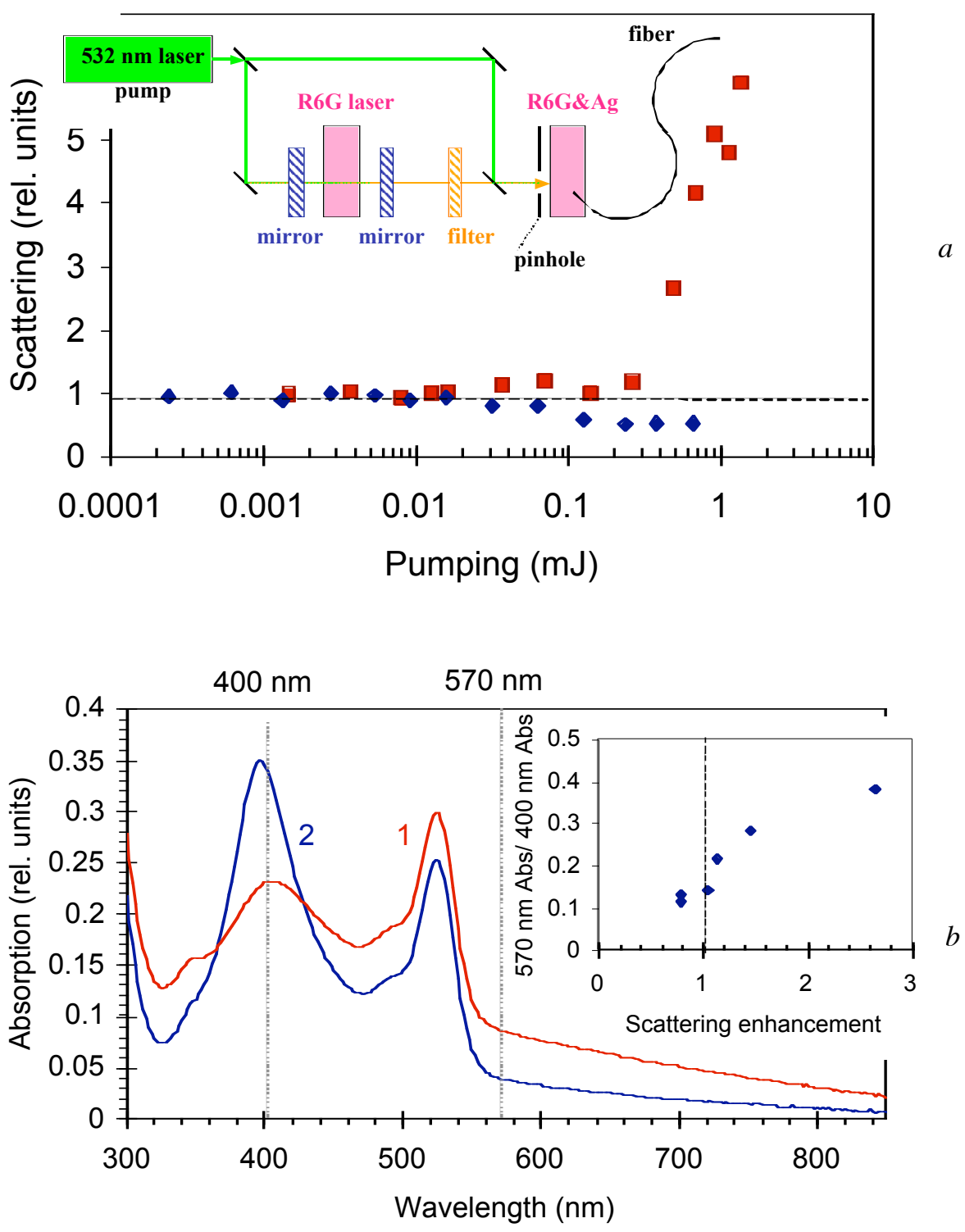

Fig. 4. (a) Intensity of the Rayleigh scattering as the function of the pumping energy. (b) Absorption spectra of the dye-Ag aggregate mixtures; R6G – $2.1\times10^{-5}$ M, Ag aggregate – $8.7\times10^{13}$ $cm^{-3}$. Squares in figure (a) and trace 1 in figure (b) correspond to one mixture, and diamonds in figure (a) and trace 2 in figure (b) correspond to another mixture. Inset of figure (a): Pump-probe experimental setup for the Rayleigh scattering measurements. Inset of figure (b): The ratio of the absorption coefficients of the dye-Ag aggregate mixtures at 570 nm and 400 nm plotted *versus* the enhancement of the Rayleigh scattering measured at 0.46 mJ.

The scattered light was collected by an optical fiber that was placed within several millimeters from the cuvette at the angles ranging from ~45° to ~135° relative the direction of the beam propagation. (We did not notice that the results of the SP enhancement measurement depended on the detection angle.) The fiber collected scattered *probe* light as well as scattered *pumping* light and spontaneous emission of dye. To separate the scattered probe light, we used a monochromator and ran the emission spectrum from 540 nm to 650 nm. The scattered probe light was seen in the spectrum as a relatively narrow (~5 nm) line on the top of a much broader spontaneous emission band.

Experimentally, we kept the energy of the probe light constant and measured the intensity of the scattered probe light as the function of the varied pumping light energy. The six-fold increase of the Rayleigh scattering observed in the dye-Ag aggregate mixture with the increase of the pumping energy (Fig. 4a, squares) is the clear experimental demonstration of the compensation of loss in metal and enhancement of the quality factor of SP resonance by the optical gain in surrounding dielectric.

The dye-Ag aggregate mixtures were placed in 1 mm thick cuvettes. At the dye concentration $2.1\times10^{-5}$ M, which we used in the majority of our experiments, the maximal optical amplification at 1 mm length was equal to ~7%. (The lateral dimension of the pumped volume was smaller than 1 mm.) The dye-Ag aggregate mixtures were visually clear, with the transport mean free path of the order of centimeters. Correspondingly, the probability of elongation of photon path in the *pumped* volume due to scattering was insignificantly small. Thus, we conclude that an increase of the intensity of scattered light in our experiment was due to an enhancement of the Rayleigh scattering cross section of metallic particles rather than simple amplification of scattered light in a medium with gain. Experimentally, no noticeable enhancement of scattering was observed in the pure R6G dye solution or pure Ag aggregate suspension.

We have found that depending on the shape of the absorption spectrum of the Ag aggregate-dye mixture (Fig. 4b) the intensity of the Rayleigh scattering could increase or decrease with the increase of pumping (Fig. 4a). Inset of Fig. 4b shows a monotonic dependence of the scattering enhancement measured at 0.46 mJ pumping energy *versus* the ratio of the absorption coefficients of the mixture at 570 and 400 nm. One can see that the relatively strong absorption of the mixture at 570 nm, which is a signature of *aggregated* Ag nanoparticles, helps to observe an enhanced Rayleigh scattering. Although we do not precisely understand the relationship between the absorption spectra and the physical properties of the mixtures, which govern the results of the scattering experiments, the correlation exists with no doubts.

A possible explanation for different scattering properties of different mixtures could be in line with the theoretical model developed in recent Ref. [31], in which a three-component system consisting of (*i*) metallic nanoparticle, (*ii*) shell of adsorbed onto it molecules with optical gain, and (*iii*) surrounding dielectric (solvent) has been studied. In particular, it has been shown that depending on the thickness of the layer of an amplifying shell, the absorption of the complex can increase or decrease with the increase of the gain in dye [31]. Similarly, we can speculate that in our experiments, in different mixtures we had different numbers of adsorbed molecules per metallic nanoparticle, which determined the enhancement or the suppression of the Rayleigh scattering.

The detailed study, which is outside the scope of this work, is required to determine the correlation between the degree of agglomeration of R6G molecules and Au nanoparticles and the absorption spectra of the mixtures. We also do not discuss in this Letter a possible relation between the compensation of SP loss by optical gain in dielectric and recently proposed surface plasmon amplification by stimulated emission of radiation (SPASER) [32], which would be a subject of separate investigation.

To summarize, we proposed the possibility of suppression of the surface resonance in metallic nanostructure embedded in a dielectric medium with absorption, the phenomenon, which is a counterpart of the enhancement of SP resonance in metallic nanoparticles submersed in a dielectric with optical gain [21]. We have observed both effects, enhancement and suppression of the SP resonance, in the mixtures of Ag aggregates and R6G dye. The demonstrated six-fold enhancement of the quality factor of the SP resonance paves the road for numerous applications of nanoplasmonics, which currently suffer from strong damping caused by absorption loss in metal.

This work was supported by the NASA grant NCC-3-1035 and the NSF grant HRD-0317722. The authors cordially thank Thomas Klar for reviewing the manuscript and discussing the results.